\documentclass[aps,prd,reprint,nofootinbib,showpacs,amsmath,amssymb]{revtex4-1}

\newcommand{\met}{m_\eta}

\newcommand{\mpn}{m_{\pi^0}}

\DeclareMathOperator{\re}{Re}

\allowdisplaybreaks[3]

\begin{document}

\title{Comment on `Determination of light quark masses from $\eta\to3\pi^0$'}

\author{Martin Zdr{\'a}hal}
 \email{zdrahal@ipnp.troja.mff.cuni.cz}
\affiliation{Institute of Particle and Nuclear Physics, Faculty of Mathematics and Physics, Charles University,
V Hole\v{s}ovi\v{c}k\'ach 2, Prague, Czech Republic}

\date{\today}

\begin{abstract}
In the commented paper [PRD 78, 034032 (2008); \href{http://arxiv.org/abs/0803.2956}{arXiv:0803.2956}] Deandrea et al.~claim that they provide a model-independent determination of the quantity $B_0(m_d-m_u)$ based on the analysis of the decay of eta to three neutral pions that takes the strong-interaction correction as an unknown parameter. In the following we show that the two constraints presented there as independent equations are in fact two approximations of a single constraint connecting together $B_0(m_d-m_u)$ with the strong-interaction correction to the $\eta\to3\pi^0$ decay amplitude. Thus, without some additional information on the latter (such as its correct value at least at one kinematic point) it does not lead to any determination of $B_0(m_d-m_u)$ and the numerical values presented there should not be taken seriously. We have also found that the numerical result for the studied quantity obtained in the commented paper is just an accidental number stemming from some numerical errors that had occurred in the computations presented there.
\end{abstract}


\maketitle

\section{Introduction}
The Bose symmetry of the decay amplitude of the $\eta\to3\pi^0$ decay translates into a very simple form of the amplitude that is valid up to the two-loop order\footnote{For a more complete discussion of this amplitude, we refer to \cite{Deandrea:2008px} and \cite{eta-phenom}.}
\begin{equation}\label{symetrie}
   \mathcal{M}(s,t,u)=M(s)+M(t)+M(u).
\end{equation}
The Mandelstam variables $s$, $t$, and $u$ fulfill the constraint
\begin{equation}\label{soucet stu}
s+t+u=\met^2+3\mpn^2\equiv3s_0.
\end{equation}

This decay violates the isospin symmetry. There are two sources of the isospin breaking, the $m_d-m_u$ mass difference and the difference of the electric charges of these two quarks. The considered decay amplitude thus has two types of contributions, the strong one proportional to $m_d-m_u$, and the electromagnetic contribution proportional to $e^2$. Neglecting the contributions of orders $(m_d-m_u)^2$, $e^4$ [and all higher ones], which should be suppressed with respect to those of the lower orders, we can write the s-wave amplitude in the form\footnote{Note that this form can be kept valid even if we include the neglected higher orders by redefinition of the meaning of the individual $\delta$s.}
\begin{equation}\label{rozklad Ms}
   M(s)=-\frac{\chi}{3\sqrt{3}F_\pi^2}\left(1+\delta_\mathrm{str}(s)+\delta_\mathrm{em}(s)\right)+\tilde{\delta}_\mathrm{em}(s).
\end{equation}
This parametrization employs the low-energy constants of the lowest order of chiral perturbation theory, the pion decay constant $F_\pi$ and the scalar quark condensate $B_0$ hidden in 
\begin{equation}
\chi\equiv B_0(m_d-m_u).   
\end{equation}
The naive power-counting $\chi\delta_\mathrm{str}(s)=O(p^2(m_d-m_u))$, $\chi\delta_\mathrm{em}(s)=O(e^2(m_d-m_u))$, $\tilde{\delta}_\mathrm{em}(s)=O(p^2e^2)$, together with the explicit computation in the first orders in chiral perturbation theory leads us to expect that the electromagnetic contributions are tiny with respect to the strong one, namely
\begin{equation}\label{potlaceni em}
\tilde{\delta}_\mathrm{em}(s)\ll \chi\delta_\mathrm{str}(s)\quad \text{ and }\quad
\delta_\mathrm{em}(s)\ll \delta_\mathrm{str}(s).                                                                                                                                                                                                                                                                                                                                                                \end{equation}
In addition, the strong contributions of higher chiral orders are expected to be significantly smaller than the leading order contribution, i.e.
\begin{equation}\label{potlaceni str}
   \delta_\mathrm{str}\ll 1.
\end{equation}

This decomposition of the decay amplitude shows how its computation can be useful for a determination of the quantity $\chi=B_0(m_d-m_u)$ giving us information about the individual $m_u$ and $m_d$ quark masses. Indeed, provided we know all of the $\delta$s appearing in (\ref{rozklad Ms}), from the total decay rate of the considered process we can determine the unknown $\chi$.

An additional independent determination of $\chi$ comes from the comparison of this amplitude squared with its experimentally measured energy dependence. In the existing experimental studies this is performed by a determination of the parameters of its expansion around the center of the Dalitz plot $s=t=u=s_0$, the so-called Dalitz plot parametrization,
\begin{equation}\label{Dalitz}
   \left\lvert\mathcal{M}(s,t,u)\right\rvert^2=\left\lvert\mathcal{M}(s_0)\right\rvert^2\left(1+2\alpha z +\dots\right),
\end{equation}
where $z$ is a dimensionless parameter denoting the distance from the center of the Dalitz plot
\begin{equation}\label{z}
   z=\frac{3}{2Q}\left((s-s_0)^2+(t-s_0)^2+(u-s_0)^2\right)
\end{equation}
with
\begin{equation}
   Q=\met^2(\met-3\mpn)^2.
\end{equation}

Note that the latter determination is possible only if we fix the physical normalization of the amplitude by $\tilde{\delta}_\mathrm{em}(s)$. If we neglect this contribution, the desired quantity $\chi$ appears in the overall normalization of the amplitude (\ref{rozklad Ms}) and even a precise measurement of the round bracket in (\ref{Dalitz}) would not enable a determination of $\chi$.

\section{Keeping the strong-interaction correction as an unknown parameter}
Having these two independent methods of the determination of $\chi$, there arises the question whether it is possible to combine them in order to determine this quantity without any knowledge on the $\delta_\mathrm{s}(s)$.

If $\delta_\mathrm{s}(s)$ was a constant, we would indeed have two independent equations for two unknowns. However, $\delta_\mathrm{s}(s)$ is a nontrivial function of $s$. In each of the methods there appears a different characteristic of this function --- in the determination that uses the decay rate of $\eta\to3\pi^0$ we have the integration of $\delta_\mathrm{s}(s)$ over the physical kinematic region, whereas in the method using the Dalitz plot parameter $\alpha$, the expansion of $\delta_\mathrm{s}(s)$ around the point $s=s_0$ is used. Even if we neglect all the higher derivatives of $\delta_\mathrm{s}(s)$, in addition to searching for a correct value of the constant $\delta_\mathrm{s}(s_0)$, we would need to solve the equations also for the second derivative $\delta_\mathrm{s}''(s_0)$, whose real part is in \cite{Deandrea:2008px} hidden in $\alpha_\mathrm{s}$.

We can therefore expect that the equations presented in \cite{Deandrea:2008px} as independent equations in which there appear just $\alpha_\mathrm{s}$ and $B_0(m_d-m_u)$ are in fact dependent and thereby insufficient for the determination of both of these unknowns. This will be shown in the rest of this comment.

\subsection{The relation following from the Dalitz plot parametrization}
The first relation follows from expanding the decay amplitude $\mathcal{M}(s,t,u)$ around the center of the Dalitz plot by using (\ref{symetrie}) and
\begin{multline}
   \mathcal{M}(s,t,u)=\sum_{j=0}^\infty\frac{1}{j!}\biggl[(s-s_0)\frac{\partial}{\partial s}+(t-s_0)\frac{\partial}{\partial t}\\ +(u-s_0)\frac{\partial}{\partial u}\biggr]^j\mathcal{M}(s,t,u)\bigg\rvert_{s=t=u=s_0}.
\end{multline}
Since $\frac{\partial}{\partial s}\mathcal{M}(s,t,u)=\frac{\partial}{\partial s}M(s)$, the symmetry of the amplitude leads to vanishing of the term with the first derivative [because of (\ref{soucet stu})] and the term with the second derivative contains $z$ from (\ref{z}),
\begin{equation}
   \mathcal{M}(s,t,u)=3M(s_0)+\frac{Q}{3}z M''(s_0)+O(z^{3/2}).
\end{equation}
With $O(z^{3/2})$ we formally denote the terms of the higher than second order in $(s-s_0)$, $(t-s_0)$ or $(u-s_0)$.
From this expansion, we compute the absolute value squared of this amplitude. By employing the form (\ref{rozklad Ms}) of its s-wave part and by comparison of the result with the Dalitz plot parametrization (\ref{Dalitz}), we obtain ``the first relation"
\begin{equation}\label{prvni relace}
   \chi^2\left(\alpha_\mathrm{str}+\alpha_\mathrm{em}\right)+\chi\tilde{\alpha}_\mathrm{em}\Delta_\pi=\chi^2\alpha,
\end{equation}
where we have denoted the real parts of the second derivatives of $\delta$s according to
\begin{gather}
   \alpha_\mathrm{str}=\frac{Q}{9}\re\delta_\mathrm{str}''(s_0),\qquad
   \alpha_\mathrm{em}=\frac{Q}{9}\re\delta_\mathrm{em}''(s_0),\\
   \tilde{\alpha}_\mathrm{em}\Delta_\pi=-F_\pi^2\frac{Q}{\sqrt{3}}\re\tilde{\delta}_\mathrm{em}''(s_0)
\end{gather}
and neglected all terms of orders $O(\chi^2\delta^2)$, $O(\tilde{\delta}_\mathrm{em}^2)$, $O(\chi\delta\tilde{\delta}_\mathrm{em})$, $O(\chi^2\delta\alpha)$, and $O(\chi\tilde{\delta}_\mathrm{em}\alpha)$. [In this counting of orders, under a shortcut $O(\delta)$ we understand both different orders $O(\delta_\mathrm{str})$ and $O(\delta_\mathrm{em})$ together. We also assume that each $\delta''(s_0)$ is of the same order as $\delta(s_0)$.]

If we assume the validity of (\ref{potlaceni em}) and (\ref{potlaceni str}) and use the smallness of the experimental value of $\alpha$, it is reasonable to neglect these orders and with this relation we have come to the same conclusions as the authors of \cite{Deandrea:2008px}. The only difference is in the numerical evaluation of $\alpha_\mathrm{em}$ as will be discussed below.

\subsection{The relation that uses the decay width}
The decay rate of $\eta\to3\pi^0$ can be obtained by the integration of the amplitude $\mathcal{M}(s,t,u)$ over the whole decay kinematic region,
\begin{equation}\label{integrace}
   \Gamma=\frac16\frac{1}{256\pi^3\met^3}\int_{4\mpn^2}^{(\met-\mpn)^2}\mathrm{d}s
   \int_{t_-(s)}^{t_+(s)}\mathrm{d}t\,\lvert\mathcal{M}(s,t,u)\rvert^2
   ,
\end{equation}
where the boundaries of the $t$-integration are
\begin{equation}
   t_\pm(s)=\frac12\left(3s_0-s\pm\lambda_{\eta\pi^0}^{1/2}(s)\sigma_{\pi^0}(s)\right)
\end{equation}
and $\lambda$ and $\sigma$ are the usual K\"{a}llen triangle function and the kinematic square root, respectively (cf.~\cite{eta-phenom}).

Note the appearance of the combinatorial factor $1/3!=1/6$ connected with the presence of three indistinguishable particles in the final state. This factor was omitted in \cite{Deandrea:2008px}. In order to reproduce the results of \cite{Deandrea:2008px} as closely as possible, we include this factor into a redefinition of $\Gamma$, i.e.~from now on with $\Gamma$ we denote the experimental value of the decay rate multiplied by the factor of 6.

We can exchange the order of integrations in (\ref{integrace}) and thanks to the symmetries of the kinematic region the boundaries of the internal $s$-integration would then be $t_\pm(t)$ depending on the variable $t$ of the external integration. Therefore, by interchanging the name of the integration variables, we obtain
\begin{multline}\label{symetrie Gamma}
   \int_{4\mpn^2}^{(\met-\mpn)^2}\mathrm{d}s
   \int_{t_-(s)}^{t_+(s)}\mathrm{d}t A(s,t)\\
   =\int_{4\mpn^2}^{(\met-\mpn)^2}\mathrm{d}s
   \int_{t_-(s)}^{t_+(s)}\mathrm{d}t A(t,s)
\end{multline}
for any function $A(s,t)$. Moreover, we can perform the same changes also with the variable $u$ and thus the integration of any function $A(s,t,u)$ gives the same result as the integration of this function with any of the Mandelstam variables interchanged.

By using this fact, the decomposition (\ref{rozklad Ms}), and neglecting again the orders $O(\chi^2\delta^2)$, $O(\tilde{\delta}_\mathrm{em}^2)$ and $O(\chi\delta\tilde{\delta}_\mathrm{em})$, we obtain the decay rate in the form,
\begin{equation}\label{Gamma}
   \Gamma=\chi^2\left(\gamma_\mathrm{tree}+\gamma_\mathrm{str}+\gamma_\mathrm{em}\right)+\chi\tilde{\gamma}_\mathrm{em},
\end{equation}
where $\gamma$s are introduced in accordance to \cite{Deandrea:2008px} as the following integrals
\begin{gather}
   \gamma_\mathrm{str}=\frac2{3F_\pi^4}\mathcal{F}[\re\delta_\mathrm{str}(s)],\quad
   \gamma_\mathrm{em}=\frac2{3F_\pi^4}\mathcal{F}[\re\delta_\mathrm{em}(s)],\\
   \gamma_\mathrm{tree}=\frac1{3F_\pi^4}\mathcal{F}[1],\quad
   \tilde{\gamma}_\mathrm{em}=-\frac{6}{\sqrt{3}F_\pi^2}\mathcal{F}[\re\tilde{\delta}_\mathrm{em}(s)].
\end{gather}
The functional $\mathcal{F}$ denotes the integration
\begin{multline}
   \mathcal{F}[f(s)]=\frac{1}{256\pi^3M_\eta^3}\int_{4M_\pi^2}^{(M_\eta-M_\pi)^2}\mathrm{d}s\int_{t_-(s)}^{t_+(s)}\mathrm{d}t\,
   f(s)\\
   =\frac{1}{256\pi^3M_\eta^3}\int_{4M_\pi^2}^{(M_\eta-M_\pi)^2}\mathrm{d}s\,
   \lambda_{\eta\pi^0}^{1/2}(s)\sigma_{\pi^0}(s)\,f(s).
\end{multline}

Now, we introduce a symbol $\overline{\Gamma}$ for the decay rate of this process in the situation the amplitude was constant and everywhere equal to its physical value at the center of the Dalitz plot, $\overline{\Gamma}=\lvert\mathcal{M}(s_0)\rvert^2\mathcal{F}[1]$. In order to be more suggestive, we divide it into three parts
\begin{equation}
   \overline{\Gamma}=\chi^2\left(\overline{\gamma}_\mathrm{tree}+\overline{\gamma}_\mathrm{str}+\overline{\gamma}_\mathrm{em}\right)+\chi\overline{\tilde{\gamma}}_\mathrm{em}
\end{equation}
with
\begin{gather}
   \overline{\gamma}_\mathrm{str}=\frac2{3F_\pi^4}\re\delta_\mathrm{str}(s_0)\mathcal{F}[1],\quad
   \overline{\gamma}_\mathrm{em}=\frac2{3F_\pi^4}\re\delta_\mathrm{em}(s_0)\mathcal{F}[1],\\
   \overline{\gamma}_\mathrm{tree}=\gamma_\mathrm{tree},\quad
   \overline{\tilde{\gamma}}_\mathrm{em}=-\frac{6}{\sqrt{3}F_\pi^2}\re\tilde{\delta}_\mathrm{em}(s_0)\mathcal{F}[1].
\end{gather}
With the help of this quantity, the decay rate can be written as
\begin{multline}\label{Gamma Ms}
   \Gamma=\overline{\Gamma}+\frac{2\chi^2}{3F_\pi^4}\mathcal{F}\left[\re\left(\delta_\mathrm{str}(s)-\delta_\mathrm{str}(s_0)\right)\right]\\
   +\frac{2\chi^2}{3F_\pi^4}\mathcal{F}\left[\re\left(\delta_\mathrm{em}(s)-\delta_\mathrm{em}(s_0)\right)\right]\\
-\frac{6\chi}{\sqrt{3}F_\pi^2}\mathcal{F}\left[\re\left(\tilde{\delta}_\mathrm{em}(s)-\tilde{\delta}_\mathrm{em}(s_0)\right)\right],
\end{multline}
where we have neglected again the same orders as in (\ref{Gamma}).

On the other hand, the integration of the Dalitz plot parametrization (\ref{Dalitz}) leads to
\begin{multline}\label{Gamma Dalitz}
   \Gamma=\lvert\mathcal{M}(s_0)\rvert^2\left(\mathcal{F}[1]+\frac{9}{Q}\alpha \mathcal{F}[(s-s_0)^2]+\mathcal{F} [O(z^{3/2})]\right)\\
   =\overline{\Gamma}+\frac{9}{Q}\alpha\lvert\mathcal{M}(s_0)\rvert^2\mathcal{F}[(s-s_0)^2]+\mathcal{F}[O(z^{3/2})].
\end{multline}

If we set the higher Dalitz plot parameters included in the dots of (\ref{Dalitz}) equal to zero or if we find $\mathcal{F} [O(z^{3/2})]$ negligible by any other reasoning, the comparison of equations (\ref{Gamma Ms}) and (\ref{Gamma Dalitz}) gives the relation
\begin{equation}\label{druha relace}
   \chi^2a_\mathrm{str}
   +\chi^2a_\mathrm{em}
-\chi\tilde{a}_\mathrm{em}(s)=-a_\mathrm{exp},
\end{equation}
where we have introduced the symbols
\begin{align}\label{astr}
   a_\mathrm{str}&=\frac{2Q}{9}\frac{\mathcal{F}\left[\re\left(\delta_\mathrm{str}(s)-\delta_\mathrm{str}(s_0)\right)\right]}{\mathcal{F}\left[(s-s_0)^2\right]}\,,\\
   a_\mathrm{em}&=\frac{2Q}{9}\frac{\mathcal{F}\left[\re\left(\delta_\mathrm{em}(s)-\delta_\mathrm{em}(s_0)\right)\right]}{\mathcal{F}\left[(s-s_0)^2\right]}\,,\\
   \tilde{a}_\mathrm{em}&=-\frac{2Q F_\pi^2}{\sqrt{3}}\frac{\mathcal{F}\left[\re\left(\tilde{\delta}_\mathrm{em}(s)-\tilde{\delta}_\mathrm{em}(s_0)\right)\right]}{\mathcal{F}\left[(s-s_0)^2\right]}\,,\label{tilde aem}\\
   a_\mathrm{exp}&=-{3F_\pi^4}\alpha\lvert\mathcal{M}(s_0)\rvert^2.\label{definice aexp}
\end{align}
From the first line of (\ref{Gamma Dalitz}), we can express the value of the amplitude at the center of the Dalitz plot and obtain
\begin{equation}\label{experimentalni aexp}
   a_\mathrm{exp}=-{3F_\pi^4}\alpha\,\frac{Q\Gamma}{Q\mathcal{F}[1]+9\alpha\mathcal{F}[(s-s_0)^2]}\,,
\end{equation}
which is exactly its definition of \cite{Deandrea:2008px}. Up to the identification of $a_\mathrm{str}=\alpha_\mathrm{str}$ performed in \cite{Deandrea:2008px}, in (\ref{druha relace}) we have obtained the ``second independent equation'' for $\chi$.

We should recall that this relation is an approximate one. Similarly, as in the first relation we have neglected the terms of orders $O(\chi^2\delta^2)$, $O(\tilde{\delta}_\mathrm{em}^2)$, $O(\chi\delta\tilde{\delta}_\mathrm{em})$. But in this relation we have also made an assumption about the higher order terms of the Dalitz plot parametrization. Setting them to zero is equivalent to setting the third and all the higher derivatives of the amplitude equal to zero. We can also understand this assumption as neglecting of the terms stemming from such higher derivatives.

This can be done also for the individual contributions from the decomposition (\ref{rozklad Ms}), i.e.~we can expand all of the $\delta$s around the center of the Dalitz plot,
\begin{equation}\label{rozvoj delta}
   \delta(s)=\delta(s_0)+(s-s_0)\delta'(s_0)+\frac12(s-s_0)^2\delta''(s_0)+\dots\,,
\end{equation}
and neglect all the terms appearing within the dots.

Thanks to the symmetries of the integration over the decay kinematic region [discussed around (\ref{symetrie Gamma})], the first derivative does not contribute,
\begin{equation}
   \mathcal{F}[s-s_0]=\frac13\mathcal{F}\left[(s-s_0)+(t-s_0)+(u-s_0)\right]=0,
\end{equation}
and from expansion (\ref{rozvoj delta}) there follows
\begin{equation}
   \mathcal{F}\left[\delta(s)-\delta(s_0)\right]=\frac12\delta''(s_0)\mathcal{F}[(s-s_0)^2]+\dots\,.
\end{equation}

In other words, if we neglect these higher-order-deri\-vative terms, we get
\begin{equation}\label{zavislost parametru}
   a_\mathrm{str}=\alpha_\mathrm{str}, \quad
   a_\mathrm{em}=\alpha_\mathrm{em}, \quad
   \tilde{a}_\mathrm{em}=\tilde{\alpha}_\mathrm{em}\Delta_\pi\,.
\end{equation}
As we have discussed above, in \cite{Deandrea:2008px} the first of these equations was used. However, if we neglect the third-order derivative of $\delta_\mathrm{str}(s)$, there is no reason for keeping the third-order derivatives of $\delta_\mathrm{em}(s)$ and $\tilde{\delta}_\mathrm{em}(s)$, as these functions should be considerably smaller than $\delta_\mathrm{str}(s)$.

With that we have identified the left-hand sides of both of the relations (\ref{prvni relace}) and (\ref{druha relace}), so we already see that they are not independent. Now, let us proceed to the right-hand sides. We start with the definition of $a_\mathrm{exp}$ (\ref{definice aexp}) and instead of evaluating $\lvert\mathcal{M}(s_0)\rvert^2$ from the experimental $\Gamma$ through (\ref{Gamma Dalitz}), we use once again the decomposition (\ref{rozklad Ms}) [which is in (\ref{druha relace}) used anyway]. By that we obtain\footnote{Note that thanks to the smallness of $\alpha$ and expecting the validity of (\ref{potlaceni em}) and (\ref{potlaceni str}), we can use here [similarly as in relation (\ref{prvni relace})] just the leading order of $M(s)$ with all $\delta$s from decomposition (\ref{rozklad Ms}) neglected. Exactly this is how we have got rid of the unknown $\delta_\mathrm{str}(s_0)$ --- however, by that we have also got rid of the genuine second independent relation [dependent on this $\delta_\mathrm{
str}(s_0)$] that was discussed in the introduction.}
\begin{equation}\label{priblizna aexp}
  a_\mathrm{exp}= -\chi^2\alpha+O(\alpha\chi^2\delta)+O(\alpha\chi\tilde{\delta}_\mathrm{em}).
\end{equation}
After using this identification, relation (\ref{druha relace}) goes exactly into relation (\ref{prvni relace}).

In conclusion, each of these relations is a different approximation of the same unique relation between the function $\delta_\mathrm{str}(s)$ and the quantity $\chi$; the first one using the smallness of $\alpha$, whereas the second one is neglecting higher-order derivatives of the amplitudes. These different omissions cannot lead to two independent relations for the unknown parameters since their use means that we assume the reasonableness of both of the omissions, which should be thus used together; and this leads to an unique approximate relation between the parameters.

\section{Conclusions}
We have shown that the ``independent parameters'' $a_\mathrm{em}$, $\tilde{a}_\mathrm{em}$ and $\alpha_\mathrm{em}, \tilde{\alpha}_\mathrm{em}$ should fulfill the approximate relations (\ref{zavislost parametru}) (up to the contributions of higher-order derivatives, which should be numerically small). Indeed, if we evaluate the expressions for $\delta$s from \cite{Deandrea:2008px}, we obtain an agreement between the values. 

We should therefore understand these relations in the following sense. We have a constraint between $\chi$ and all the functions $\delta(s)$, which is in the form of (\ref{druha relace}), where we have two different ways how to approximate the parameters $a_i$ from its left-hand side, either through (\ref{astr})--(\ref{tilde aem}) or through (\ref{zavislost parametru}). From these two approximations, we can determine the values of $a_i$ and estimate their errors stemming from the higher-order corrections neglected in these two different enumerations. 
Similarly, the right-hand side of this relation, the parameter $a_\mathrm{exp}$, can be obtained\footnote{According to the author of this comment, the determination 
(\ref{experimentalni aexp}) (with the correct definition of $\Gamma$ including the factor of 6) is more reasonable
as it ignores just an integral of higher Dalitz plot parameters, which should be small due to the smallness of the decay kinematic region of this process. In fact, after a measurement of the higher Dalitz parameters, one can simply make its value more precise by adding these parameters into (\ref{Gamma Dalitz}). On the other hand, the determination (\ref{priblizna aexp}) rests heavily on the condition that $\delta$s in (\ref{rozklad Ms}) are very small. However, the comparison of these two determination by using the value of $B_0$ from the recent lattice determinations and of $m_d-m_u$ from \cite{eta-phenom} leads to $\sqrt{\frac{a_\mathrm{exp}}{-\alpha\chi^2}}\approx1.9$, i.e.~$\delta$s can be significantly larger than expected and one should be careful when employing (\ref{potlaceni em}) and (\ref{potlaceni str}).} either from (\ref{experimentalni aexp}) or from (\ref{priblizna aexp}).
In order to solve this constraint and obtain the quantity $B_0(m_d-m_u)$ from it, we need some additional assumption or constraint on the physical amplitude $\mathcal{M}(s,t,u)$ or on its part $\delta_\mathrm{str}(s)$.
An example of such an additional condition is to set the value of $\delta_\mathrm{str}(s)$ at some specific point, where we for instance believe its chiral computation (cf.\ \cite{eta-phenom}).

Why have the authors of \cite{Deandrea:2008px} not found these conclusions in their numerical analysis? We have checked the results of \cite{Deandrea:2008px} and found an error in the expression given there for $\alpha_\mathrm{em}^{(1)}$, which does give the (wrong) numerical value of this quantity that has been presented in \cite{Deandrea:2008px}. Similarly, the numerical values of $\alpha$, $a_\mathrm{exp}$ and the value of $\chi$ should approximately [cf.~footnote 4] fulfill relation (\ref{priblizna aexp}), which is not the case for the values presented in \cite{Deandrea:2008px}. As we have discussed above, the reason is that in the relation (\ref{experimentalni aexp}) one should introduce not the experimental value of $\Gamma$ but this value multiplied by 6. Both of these errors shift the coefficients of relation (\ref{druha relace}) by a 
quantity of the order of the correct coefficients, and such a shift comprises an unphysical second independent constraint on $\alpha_\mathrm{str}$ and $\chi$ in addition to the unique independent approximate relation (\ref{druha relace}). This explains quite a reasonable magnitude of the values for $\chi=B_0(m_d-m_u)$ and $\alpha_\mathrm{str}$ obtained from such an analysis in \cite{Deandrea:2008px}. We stress once again that these numbers are chosen from the whole two-parametrical space of solutions of relation (\ref{druha relace}) just by accident and one should not take them seriously.

\acknowledgments{%
The author is grateful to J.~Ho\v{r}ej\v{s}\'{i} and J.~Novotn\'{y} for discussions on this text. This work is supported by the Ministry of Education, Youth and Sports of the Czech Republic (MSM0021620859) and by the Charles University in Prague (UNCE 204020/2012).}

\bibliography{bibliografie}

\begin{thebibliography}{2}%
\makeatletter
\providecommand \@ifxundefined [1]{%
 \@ifx{#1\undefined}
}%
\providecommand \@ifnum [1]{%
 \ifnum #1\expandafter \@firstoftwo
 \else \expandafter \@secondoftwo
 \fi
}%
\providecommand \@ifx [1]{%
 \ifx #1\expandafter \@firstoftwo
 \else \expandafter \@secondoftwo
 \fi
}%
\providecommand \natexlab [1]{#1}%
\providecommand \enquote  [1]{``#1''}%
\providecommand \bibnamefont  [1]{#1}%
\providecommand \bibfnamefont [1]{#1}%
\providecommand \citenamefont [1]{#1}%
\providecommand \href@noop [0]{\@secondoftwo}%
\providecommand \href [0]{\begingroup \@sanitize@url \@href}%
\providecommand \@href[1]{\@@startlink{#1}\@@href}%
\providecommand \@@href[1]{\endgroup#1\@@endlink}%
\providecommand \@sanitize@url [0]{\catcode `\\12\catcode `\$12\catcode
  `\&12\catcode `\#12\catcode `\^12\catcode `\_12\catcode `\%12\relax}%
\providecommand \@@startlink[1]{}%
\providecommand \@@endlink[0]{}%
\providecommand \url  [0]{\begingroup\@sanitize@url \@url }%
\providecommand \@url [1]{\endgroup\@href {#1}{\urlprefix }}%
\providecommand \urlprefix  [0]{URL }%
\providecommand \Eprint [0]{\href }%
\providecommand \doibase [0]{http://dx.doi.org/}%
\providecommand \selectlanguage [0]{\@gobble}%
\providecommand \bibinfo  [0]{\@secondoftwo}%
\providecommand \bibfield  [0]{\@secondoftwo}%
\providecommand \translation [1]{[#1]}%
\providecommand \BibitemOpen [0]{}%
\providecommand \bibitemStop [0]{}%
\providecommand \bibitemNoStop [0]{.\EOS\space}%
\providecommand \EOS [0]{\spacefactor3000\relax}%
\providecommand \BibitemShut  [1]{\csname bibitem#1\endcsname}%
\let\auto@bib@innerbib\@empty
\bibitem [{\citenamefont {Deandrea}\ \emph {et~al.}(2008)\citenamefont
  {Deandrea}, \citenamefont {Nehme},\ and\ \citenamefont
  {Talavera}}]{Deandrea:2008px}%
  \BibitemOpen
  \bibfield  {author} {\bibinfo {author} {\bibfnamefont {A.}~\bibnamefont
  {Deandrea}}, \bibinfo {author} {\bibfnamefont {A.}~\bibnamefont {Nehme}}, \
  and\ \bibinfo {author} {\bibfnamefont {P.}~\bibnamefont {Talavera}},\ } {\bibfield  {journal} {\bibinfo
  {journal} {Phys.\ Rev.}\ }\textbf {\bibinfo {volume} {D78}},\ \bibinfo
  {pages} {034032} (\bibinfo {year} {2008})},\ \Eprint
  {http://arxiv.org/abs/0803.2956} {arXiv:0803.2956 [hep-ph]} \BibitemShut
  {NoStop}%
\bibitem [{\citenamefont {Kampf}\ \emph {et~al.}(2011)\citenamefont {Kampf},
  \citenamefont {Knecht}, \citenamefont {Novotn{\'y}},\ and\ \citenamefont
  {Zdr{\'a}hal}}]{eta-phenom}%
  \BibitemOpen
  \bibfield  {author} {\bibinfo {author} {\bibfnamefont {K.}~\bibnamefont
  {Kampf}}, \bibinfo {author} {\bibfnamefont {M.}~\bibnamefont {Knecht}},
  \bibinfo {author} {\bibfnamefont {J.}~\bibnamefont {Novotn{\'y}}}, \ and\
  \bibinfo {author} {\bibfnamefont {M.}~\bibnamefont {Zdr{\'a}hal}},\ } {\bibfield  {journal} {\bibinfo
  {journal} {Phys.\ Rev.}\ }\textbf {\bibinfo {volume} {D84}},\ \bibinfo
  {pages} {114015} (\bibinfo {year} {2011})},\ \Eprint
  {http://arxiv.org/abs/1103.0982} {arXiv:1103.0982 [hep-ph]} \BibitemShut
  {NoStop}%
\end{thebibliography}%

\end{document}